\documentclass[a4paper,12pt]{article}
\pdfoutput=1

\usepackage{amsmath,amssymb,amsfonts,mathtools} 
\usepackage{physics}
\allowdisplaybreaks
\usepackage{graphicx} 
\usepackage{color} 
\usepackage{caption}
\usepackage{subcaption}
\usepackage[bottom]{footmisc}	
\makeatletter
\@addtoreset{equation}{section}
\renewcommand{\theequation}{\thesection.\@arabic\c@equation}
\makeatother
\usepackage{hyperref} 
    \hypersetup{colorlinks=true, linktocpage, linkcolor={blue}, citecolor={blue}, urlcolor={blue}}  
\usepackage{cite}

\def \be {\begin{equation}}
\def \ee {\end{equation}}
\def \nn {\nonumber}

\def \del {\partial}
\def \Lam {\Lambda}
\def \om {\omega}

\def \vs {v_{\mathrm{s}}}		
\def \chiOm {\chi_{{\textstyle\mathstrut}\Omega}}    
\def \aOm {a_{\vphantom{1^{H}}\Omega}} 
\def \chiOmc {\chi^*_{\vphantom{1^{H}}\Omega}}
\def \chiom {\chi_{{\textstyle\mathstrut}\omega}}

\begin{document}

\begin{titlepage}
\vspace*{-10mm}   
\baselineskip 10pt   
\begin{flushright}   
\begin{tabular}{r} 
\end{tabular}   
\end{flushright}   
\baselineskip 24pt   
\vglue 10mm

\begin{center}

\noindent
\textbf{\LARGE
UV Dispersive Effects on Hawking Radiation
\\
\vskip0.3em
}
\vskip 10mm
\baselineskip 20pt

\renewcommand{\thefootnote}{\fnsymbol{footnote}}

{\large
Emil T. Akhmedov${}^{a,b}$
\footnote[1]{akhmedov@itep.ru},
Tin-Long Chau${}^{c}$
\footnote[2]{ronnychau031@gmail.com},
Pei-Ming~Ho${}^{c,d}$
\footnote[3]{pmho@phys.ntu.edu.tw},
Hikaru~Kawai${}^{c,d}$
\footnote[4]{hikarukawai@phys.ntu.edu.tw},
Wei-Hsiang Shao${}^{c}$
\footnote[5]{whsshao@gmail.com},
and Cheng-Tsung Wang${}^{c}$
\footnote[6]{ctgigglewang@gmail.com}
}

\renewcommand{\thefootnote}{\arabic{footnote}}

\vskip5mm

{\it
${}^{a}$
Moscow Institute of Physics and Technology, 
141700, Dolgoprudny, Russia \\
${}^{b}$
National Research Centre ``Kurchatov Institute'', 123182, Moscow, Russia \\
${}^{c}$
Department of Physics and Center for Theoretical Physics, \\
National Taiwan University, Taipei 10617, Taiwan \\
${}^{d}$
Physics Division, National Center for Theoretical Sciences, Taipei 10617, Taiwan
}

\vskip 15mm
\begin{abstract}
We revisit the connection between Hawking radiation 
and high-frequency dispersions for a Schwarzschild black hole
following the work of Brout et al.~\cite{Brout:1995wp}. 
After confirming the robustness of Hawking radiation
for monotonic dispersion relations,
we consider non-monotonic dispersion relations 
that deviate from the standard relation only in the trans-Planckian domain.
Contrary to the common belief that Hawking radiation is insensitive to UV physics,
it turns out that Hawking radiation is subject to significant modifications
after the scrambling time.
Depending on the UV physics at the singularity,
the amplitude of Hawking radiation 
could diminish after the scrambling time,
while the Hawking temperature remains the same.
Our finding is thus not contradictory to earlier works 
regarding the robustness of Hawking temperature.
\end{abstract}
\end{center}

\end{titlepage}

\pagestyle{plain}

\baselineskip 18pt

\setcounter{page}{1}
\setcounter{footnote}{0}
\setcounter{section}{0}

\tableofcontents


\section{Introduction}
\label{sec:intro}

It is a common folklore that Hawking radiation~\cite{Hawking:1974rv, Hawking:1975vcx} is 
a robust prediction about black holes,
despite the fact that
its derivation based on effective theories
actually involves trans-Planckian modes~\cite{tHooft:1984kcu, Jacobson:1991gr}.
Naively,
one would expect that different UV physics lead to different properties of Hawking radiation.
Yet, 
in a series of works~\cite{Unruh:1994je, Brout:1995wp, Corley:1996ar, Corley:1997pr, Himemoto:1999kd, Unruh:2004zk, Schutzhold:2008tx, Coutant:2011in, Schutzhold:2013mba, Coutant:2014cwa},
various modifications to the dispersion relation 
of the radiation field 
were considered in the trans-Planckian regime $p \gtrsim M_p$ (where $M_p$ stands for the Planck mass),
and approximately the same Hawking radiation
was reproduced repeatedly.
These works led some to believe that Hawking radiation is a robust prediction.
It also led to the developments of the subject called ``analogue gravity''~\cite{Barcelo:2005fc}.
Strictly speaking,
Hawking radiation is a transient phenomenon.
When the radius of the spherical collapsing matter is 
still much larger than the Schwarzschild radius,
there is of course no Hawking radiation.
Hawking radiation starts to appear when the distance between
the surface of the collapsing matter and the Schwarzschild radius
is much smaller than the Schwarachild radius.
After that,
it is often assumed that the evaporation of a black hole is an adiabatic process,
and Hawking radiation persists until the black hole's mass becomes
of the order of the Planck scale.
However,
it was pointed out in Ref.~\cite{Akhmedov:2015xwa} that
two-loop corrections to the free field theory
exhibit secular growth that leads to the breakdown of the perturbation theory.
It was also found~\cite{Ho:2021sbi, Ho:2022gpg} that,
if higher-derivative interactions between the radiation field
and the background geometry are taken into consideration,
Hawking radiation can be significantly modified after the scrambling time.
(For a black hole with Schwarzschild radius $a$,
the scrambling time is $\mathcal{O}(2a\log(aM_p))$~\cite{Sekino:2008he}.)
We are thus motivated to revisit the connection between
Hawking radiation and high-frequency dispersions.
We adopt the approach of Brout et al.~\cite{Brout:1995wp} 
and consider both monotonic and non-monotonic dispersion relations.
Special attention is paid to 
the time-dependence of the magnitude of Hawking radiation,
while previous works on this topic 
have mostly focused on the Hawking temperature.
Since the magnitude can change without changing the temperature,
our study may reveal new features of Hawking radiation that were missed in the past.
It turns out that,
for certain non-monotonic dispersion relations,
Hawking radiation can become significantly different.
For instance,
the magnitude of Hawking radiation can decay exponentially with time,
so that it is effectively turned off after a certain critical time.
Depending on the dispersion relation,
this critical time can be as short as the scrambling time.
In Sec.~\ref{sec:BMPS},
we review the formulation of Ref.~\cite{Brout:1995wp}
to investigate the relation between Hawking radiation and high-frequency dispersions.
In Sec.~\ref{sec:monotonic},
we consider generic monotonic dispersion relations
that are subluminal outside the horizon,
and observe how Hawking radiation is independent of 
such UV modifications of monotonic dispersion relations.
In Sec.~\ref{sec:nonmonotonic},
we consider generic non-monotonic dispersion relations,
and find that,
when the dispersion curve $g(p)$
approaches zero faster than $\mathcal{O}(1/p)$ in the large momentum limit $p \rightarrow \infty$,
Hawking radiation is modified after a critical time 
depending on $g(p)$.
The physics behind this is explained in the same section,
and we conclude in Sec.~\ref{sec:conclude}.

\section{The Setup}
\label{sec:BMPS}

We follow the work of Brout, Massar, Parentani and Spindel~\cite{Brout:1995wp} 
and study the evolution of Hawking quanta on a spherically-symmetric black-hole background
with modified dispersion relations defined with respect to the ingoing Vaidya metric.
Due to its simplicity and convenience,
similar settings with the same coordinate system were adopted in related works 
such as Refs.~\cite{Damour:1976jd, Brout:1998ei, Ren:2006zu, Greenwood:2008zg, Abdolrahimi:2016emo}, 
in which the leading behavior of Hawking radiation was confirmed. 
\begin{figure}[ht]
\centering
\includegraphics[scale=0.7]{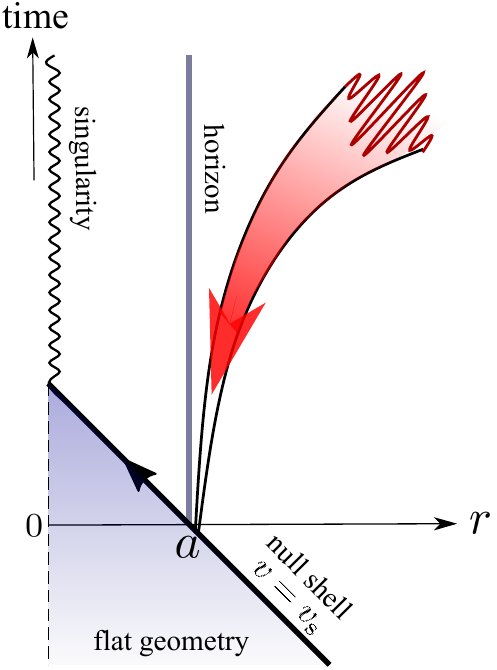}
\caption{A thin null shell with flat Minkowski interior collapses to form a black hole, 
displayed in the ingoing Eddington-Finkelstein coordinates. 
We trace an outgoing wave packet backwards in time and
decompose it into Minkowski modes inside the shell.}
\label{fig:vaidya}
\end{figure}
For simplicity,
the collapsing matter is assumed to be a spherically-symmetric thin null shell 
falling along the trajectory $v = \vs$ (see Fig.~\ref{fig:vaidya}).
The Schwarzschild metric outside the shell is connected 
with the interior Minkowski geometry along the shell, and the corresponding ingoing Vaidya metric reads
\begin{equation}
\label{eq:vaidya}
ds^2 = -\left( 1 - \Theta(v - \vs) \, \frac{a}{r} \right)dv^2 + 2dvdr + r^2 d\Omega^2 \, ,
\end{equation}
where $a$ is the Schwarzschild radius.
Without loss of generality,
we can take
\be
\vs = 0 \, .
\ee
We will neglect the angular dependence from this point onward 
and consider only $s$-waves on this background.
As for the radiation field, we consider a massless real scalar described by the action
\be 
\label{eq:action}
S = -\frac{1}{2} \, \int dr \int dv \, \left[ g(i \del_r) \phi \right] \left[ -2 i \del_v + \left( 1 - \Theta(v) \, \frac{a}{r} \right) g(-i \del_r) \right] \phi \, ,
\ee
where the UV dispersion is introduced through the pseudo-differential operator $g(-i\partial_r)$.
The function $g(p)$ is assumed to be odd so that the action is Hermitian.
We have $g(p) = p$ for the standard dispersion relation of a massless field.
Variation of the action leads to the field equation
\be 
\label{eq:wave_eqn}
g(-i \del_r) 
\left[ -2 i \del_v + \left( 1 - \Theta(v) \, \frac{a}{r} \right) g(-i \del_r) \right] \phi (v, r) 
= 0 \, .
\ee
The junction condition for $\phi$ at $v = \vs$ is simply its continuity. 
General solutions of eq.~\eqref{eq:wave_eqn} are superpositions of the ingoing and outgoing modes.
The ingoing modes $\phi_{\mathrm{in}} = \phi_{\mathrm{in}}(v)$ satisfy $g(-i\partial_r) \phi_{\mathrm{in}}(v, r) = 0$, whereas the outgoing modes $\phi_{\mathrm{out}}(v, r)$ satisfy
\begin{equation}
\label{eq:wave_eqn_out}
\left[ -2 i \del_v + \left( 1 - \Theta(v) \, \frac{a}{r} \right) g(-i \del_r) \right] \phi_{\mathrm{out}}(v, r) = 0 \, .
\end{equation}
As the ingoing and outgoing modes are decoupled, 
we shall focus solely on the outgoing sector, 
which is where Hawking radiation resides,
and simply denote $\phi_{\mathrm{out}}$ as $\phi$.
The dispersion curve $g(p)$ encodes how the propagation of outgoing modes 
is deformed by new physics above the Planck scale $M_p$.
In the IR regime $p \ll M_p$,
it reduces to $g(p) \approx p$, 
and the standard low-energy effective theory is recovered. 
In the black hole spacetime outside the shell (for $v > 0$), 
eq.~\eqref{eq:wave_eqn_out} results in the space-dependent dispersion relation
\be 
\label{eq:disprel_out}
\omega = \frac{1}{2} \left( 1 - \frac{a}{r} \right) g(p) \, .
\ee
In the flat spacetime inside the shell ($v < 0$), 
we instead have 
\be 
\Omega = \frac{g(p)}{2} \, ,
\ee
where $\Omega$ is the eigenvalue of $- i \partial_v$.
In the following,
we shall compute Hawking radiation for the free field theory
defined above for different types of dispersion relations.

\section{Monotonic Dispersion}
\label{sec:monotonic}

In this section, we generalize the discussion of Ref.~\cite{Brout:1995wp} by 
considering generic monotonic dispersion curves $g(p)$ that are \emph{subluminal} outside the horizon, 
and demonstrate that Hawking radiation is robust against such UV modifications.

\subsection{Particle description}

\begin{figure}[ht]
\centering
\includegraphics[scale=1.2]{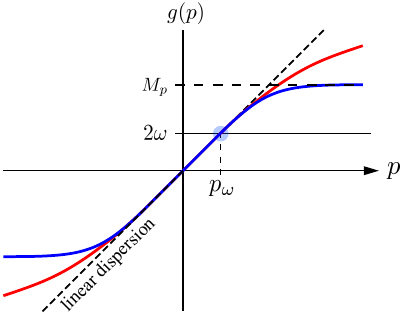}
\caption{A schematic plot illustrating two examples of monotonic dispersion curves 
for which the group velocity is subluminal outside the horizon.
These profiles deviate significantly from standard linear dispersion $g(p) = p$ 
only for trans-Planckian momenta $p \gtrsim M_p$.}
\label{fig:i}
\end{figure}

To begin with, 
the particle description of the Hawking quanta is as follows.
Treating the coordinate $v$ as the time variable,
the Hamiltonian can be read off from eq.~\eqref{eq:disprel_out} as
\be 
\label{eq:hamiltonian}
H(p, r) \equiv \frac{1}{2} \left( 1 - \frac{a}{r} \right) g(p) \, .
\ee
The corresponding Hamilton equations are
\begin{align}
    \label{eq:hamilton1}
    \frac{dr}{dv} &= \frac{\partial H}{\partial p} = \frac{1}{2} \left( 1 -\frac{a}{r} \right) g'(p) \, , \\
    \frac{dp}{dv} &= - \frac{\partial H}{\partial r} = - \frac{a}{2r^2} \, g(p) \, .
    \label{eq:hamilton2}
\end{align}
In terms of the Vaidya line element~\eqref{eq:vaidya}, 
in order for a wave packet outside the horizon to exhibit subluminal propagation, 
it is necessary that the group velocity satisfies $dr / dv < \left( 1 - a / r \right) / 2$, which implies the condition 
\be 
\label{eq:sublum}
g'(p) < 1 \, .
\ee
Fig.~\ref{fig:i} displays the types of monotonic profiles that fulfill this requirement. 
We consider generic monotonic $g(p)$ satisfying eq.~\eqref{eq:sublum} in this section.

\begin{figure}[ht]
\centering
\includegraphics[scale=1.0]{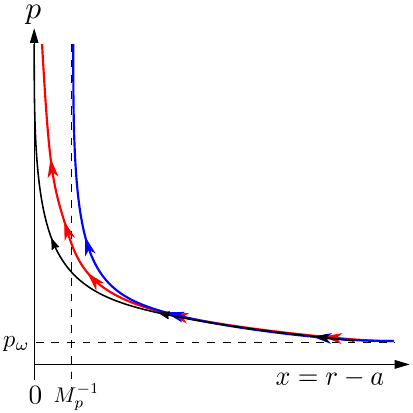}
\caption{
A sketch of the classical phase space trajectories.
The arrows indicate the direction of \emph{decreasing} time $v$.
The blue and red curves represent the trajectories
of Hawking quanta for the dispersion profiles shown in blue and red, 
respectively, in Fig.~\ref{fig:i},
and the black curve for the standard low-energy effective theory.
}
\label{fig:j}
\end{figure}

We comment that requiring subluminality~\eqref{eq:sublum} outside the horizon 
automatically gives rise to superluminal behavior inside the horizon 
due to a sign flip coming from $(1 - a / r)$ in eq.~\eqref{eq:hamilton1}. 
While this is an exclusive feature in the $(v, r)$ coordinate system, 
it should not be viewed as a pathology of the model.
As Lorentz symmetry has already been violated by modifying the dispersion relation, the speed of light defined by $ds^2 = 0$ no longer has a universal meaning\,\footnote{
For instance, superluminal modes are also present in Ho\v{r}ava-Lifshitz gravity~\cite{Horava:2009uw}
and Einstein-Aether theory~\cite{Gasperini:1987nq, Jacobson:2000xp}.}.
The condition $ds^2 < 0$ for the line element~\eqref{eq:vaidya} should be understood as the subluminality condition only in the low-energy limit, 
while the precise definition of super/subluminality is given by the wave equation of a massless field.
A Hawking particle detected at large distances 
is originated from the Minkowski vacuum inside the null shell.
To understand the emergence of Hawking radiation,
we trace a Hawking particle backward in time,
from large $r$ to the near-horizon region and beyond.
Its wave packet composed of purely positive-frequency modes at large $r$
is expected to turn into a mixture of positive and negative-frequency Minkowski modes inside the shell.
This mixture determines the magnitude and temperature of Hawking radiation.
Denote by $u$ the Eddington retarded time defined by 
\be
u \equiv v - 2r_{\ast} \, ,
\ee
where $r_{\ast}$ is the tortoise coordinate 
\be
r_{\ast} \equiv r + a\log(r/a - 1) \, .
\ee
An outgoing particle at low energy moves along a constant-$u$ trajectory at large distances.
Consider an outgoing wave packet centered around the retarded time $u = u_0$ 
near asymptotic infinity ($r \to \infty$) 
with a small but positive Killing frequency $\omega \ll M_p$. 
This wave packet has a momentum centered around 
\be 
\label{eq:pole1}
p_{\omega} \equiv g^{-1} (2 \omega) \simeq 2 \omega \, ,
\ee
where $p_{\omega}$ is the root of the dispersion relation~\eqref{eq:disprel_out} at large $r$ (see Fig.~\ref{fig:i})\,\footnote{
For monotonic $g(p)$, this root is uniquely defined. 
}.
As we propagate this wave packet backwards in time $v$, since its frequency $\omega$ is a constant of motion, its central momentum can be inferred from its position via 
eq.~\eqref{eq:disprel_out}.
As $g(p) \approx p$ is assumed to be a good approximation for $p \lesssim M_p$,
the Hawking particle follows the trajectory in the low-energy effective theory
as long as $p \lesssim M_p$.
According to eq.~\eqref{eq:disprel_out},
for a wave packet with the dominant frequency $\omega \sim 1/a$ in Hawking radiation,
the momentum $p$ increases from $p_{\omega}$~\eqref{eq:pole1} towards $M_p$ 
as we trace its trajectory backward in time.
In the near-horizon region ($|r - a| \ll a$),
it is convenient to define
\be
x \equiv r - a \, . 
\ee
We find that $p \sim M_p$ and $x \sim 1/M_p$
when $v \sim u_0 - 2a \log(a M_p)$.
The UV modification to the dispersion relation is irrelevant up to this point.
After the wave packet reaches the region $x \sim M_p^{-1}$, 
its evolution backwards in time is sensitive to the UV-modification of $g(p)$.
For $x \ll a$,
eqs.~\eqref{eq:hamilton1} and~\eqref{eq:hamilton2}
can be approximated by
\begin{align}
\label{eq:dxdv0}
\frac{dx}{dv} \approx \frac{g'(p) x}{2 a} \, ,
\\
\frac{dp}{dv} \approx - \frac{g(p)}{2a} \, .
\end{align}
They lead to
\be 
\label{eq:x_exp}
x \approx \frac{\mbox{const}}{g(p)} \, .
\ee 
Since $g'(p)$ is non-negative for monotonic $g(p)$, 
$p$ increases continuously in the $-v$ direction.
This implies that eventually (at sufficiently early times)
the wave packets are crammed against a point 
which is either the horizon ($x = 0$) or a finite distance from it 
(see Fig.~\ref{fig:j}), 
depending on whether $g(p \to \infty)$ is infinite or not.
In both cases, 
Hawking quanta are originated from outgoing modes that are exponentially compressed in the past.
This is the primary reason why Hawking radiation turns out to remain essentially the same, as we will see below.

\subsection{Robustness of Hawking radiation}
\label{sec:HR_mono}

Let us now compute the Hawking radiation for monotonic dispersion relations
that are subluminal outside the horizon. 
We will see that the effect of modifying the dispersion relation
can be undone by a change of variables.
In general, 
as $g(-i\partial_r)$ involves spatial derivatives of infinite order, 
the wave equation~\eqref{eq:wave_eqn_out} 
admits an infinite number of linearly independent solutions.
But since we are only interested in solutions corresponding to Hawking quanta, 
we will only consider solutions of propagating modes.
It is thus convenient to work in the momentum space, 
where propagating modes can be readily extracted by 
suitably deforming the integration contour around singularities 
and branch cuts~\cite{Brout:1995wp, Corley:1997pr, Himemoto:1999kd, Unruh:2004zk, Schutzhold:2008tx}. 
In the momentum space representation, the wave equation~\eqref{eq:wave_eqn_out} for outgoing modes outside the collapsing shell takes the form
\be 
\del_p \big\{ \left[ 2 i \del_v - g(p) \right] \widetilde{\phi}(v, p) \big\} = i a g(p) \, \widetilde{\phi}(v, p) \, ,
\ee 
and the stationary solutions $\widetilde{\phi}_{\omega}(v, p) = e^{-i \omega v} \, \widetilde{\varphi}_\omega(p)$ can be solved to give
\be
\label{eq:wavesoln} 
\widetilde{\varphi}_\omega(p) = \mathcal{N}_{\omega} \, \frac{e^{-ipa}}{2\omega - g(p-i0^+)} \exp\left[ 2 i a \omega \int_0^p \frac{dk}{2 \omega - g(k-i0^+)} \right] \, ,
\ee
where $\mathcal{N}_\omega$ is the normalization constant.
The $i 0^+$-prescription is chosen such that
these solutions represent Hawking modes at large $r$. 
To understand this, 
we perform an inverse Fourier transform of eq.~\eqref{eq:wavesoln} 
to examine its properties in the position space:
\be 
\phi_\omega (v, r) = 
\mathcal{N}_{\omega} \, e^{-i \omega v} \int_{-\infty}^{\infty}\frac{dp}{2\pi} \, 
\frac{e^{i p (r - a)}}{2\omega-g(p-i0^+)} \, 
\exp\left[ 2 i a \omega \int_0^p \frac{dk}{2 \omega - g(k-i0^+)} \right] \, .
\ee
The Riemann–Lebesgue lemma implies that this integral vanishes as $r \to \infty$ 
except in a small neighborhood of the pole at $p_{\omega}$. 
Therefore, in the asymptotic region, the modes can be approximated by
\begin{align}
    \phi_\omega (v, r \to \infty) &\approx \mathcal{N}_{\omega} \, 
    e^{-i \omega v} \int_{-\infty}^{\infty} \frac{dp}{2\pi} \, 
    \frac{e^{i p (r - a)}}{g_{\omega}' (p_{\omega} - p) + i0^+} \, 
    \exp\left[ 2 i a \omega \int_0^p \frac{dk}{g_{\omega}' (p_{\omega} - k) + i0^+} \right] 
    \nn \\
    &\approx
    \frac{\mathcal{N}_{\omega}}{2 \pi g_{\omega}'} 
    \big( e^{- \pi a \omega / g_{\omega}'} - e^{3 \pi a \omega / g_{\omega}'} \big) \, 
    \Gamma \left( -2i a \omega / g_{\omega}' \right) 
    e^{-i \omega v \, + \, i p_{\omega} r \, + \, \mathcal{O}(\log r)} \, ,
    \label{eq:soln_infty}
\end{align}
up to a constant phase, where 
\be 
g_{\omega}' \equiv g'(p_{\omega}) \, .
\label{eq:g_prime}
\ee
We see that $\phi_\omega(v, r \to \infty) \propto e^{-i\omega v \, + \, i p_{\omega} r}$ 
indeed reduces to single-frequency plane waves to the leading order at large distances, 
so they can be superposed to form wave packets that represent Hawking particles.
By the same token, 
it can be shown that,
with this $i 0^+$-prescription,
$\phi_\omega(v, r)$ decays rapidly inside the horizon ($r < a$)~\cite{Corley:1997pr, Himemoto:1999kd, Unruh:2004zk}, 
which is a desired property for outgoing Hawking particles\,\footnote{
The solutions~\eqref{eq:wavesoln} with the $i 0^-$-prescription 
correspond to the Hawking partners inside the horizon.
}.
The outgoing sector of the field $\phi$ outside the shell can thus be 
expanded in terms of these solutions $\phi_\omega(v, r)$ as
\be 
\label{eq:expand_b}
\phi(v > 0, r) = 
\int_0^{\Lambda} \frac{d\omega}{2\pi} \left[ 
b_\omega \, \phi_\omega(v, r) + b_\omega^\dagger \, \phi_\omega^*(v, r) \right] \, ,
\ee
where $\Lambda= g(\infty)/2$ is the upper bound of the frequency $\omega$ at large $r$
($\Lambda$ can be infinity). 
The mode expansion inside the shell can be written as
\be 
\label{eq:expand_a}
\phi(v < 0, r) = 
\int_0^{\Lambda}\frac{d\Omega}{2\pi} 
\left[ 
\aOm \, \chiOm (v, r) + a_\Omega^\dagger \, \chiOmc(v, r) 
\right] \, ,
\ee 
where 
\be 
\label{eq:chiOm}
\chiOm(v ,r) = \frac{1}{\sqrt{2\Omega \, g_{\Omega}'}} \, e^{-i\Omega v \, + \, i p_\Omega r}
\ee
is an outgoing mode in flat space
obeying the wave equation $\left[ -2 i \del_v + g(-i \del_r) \right] \phi = 0$. 
Since the Hawking modes~\eqref{eq:soln_infty} coincide with 
the flat-space basis $\chiom(v, r)$ in the asymptotically flat region $r \to \infty$, 
the normalization constant $\mathcal{N}_\omega$ in eq.~\eqref{eq:wavesoln} 
can be fixed as
\be 
\label{eq:normalization}
\mathcal{N}_\omega = \sqrt{\frac{2\pi a}{e^{4\pi a \omega / g_{\omega}'}-1}} \, .
\ee 
${\cal N}_{\om}$ is insensitive to the UV-modification of $g(p)$ for $\om \ll M_p$.
As mentioned in Sec.~\ref{sec:BMPS}, 
the field has to be continuous across the null shell. 
Imposing this matching condition between the mode expansions inside and outside the shell relates the two sets of coefficients $\aOm$ and $b_{\omega}$ via the Bogoliubov transformation
\be 
\label{eq:bogoliubov}
b_\omega = \int_0^\Lambda\frac{d\Omega}{2\pi} \left[ \alpha_{\omega\Omega} \, \aOm + \beta_{\omega\Omega} \, a_\Omega^\dagger \right] \, ,
\ee 
where
\be 
\label{eq:alphabeta}
\alpha_{\omega \Omega} = 2\sqrt{\frac{2\Omega}{g_{\Omega}'}} \, \widetilde{\varphi}_{\omega}^{\,*}(p_\Omega) \, ,
\qquad 
\beta_{\omega \Omega} =-2\sqrt{\frac{2\Omega}{g_{\Omega}'}} \, \widetilde{\varphi}_{\omega}^{\,*}(-p_\Omega) 
\ee
can be obtained through the Fourier transform of $\varphi_\omega(r)$. We refer the reader to appendix~\ref{app:bogoliubov} for a detailed derivation. 
To quantize the field, $\phi$ and its conjugate momentum $\Pi = -ig(-i\partial_r)\phi$ are promoted to operators satisfying the equal-time commutation relation $\comm{\phi(v, x)}{\Pi(v, y)} = i \delta(x - y)$.
This leads to the canonical commutation relations 
\be 
\label{eq:comm}
\comm*{\aOm}{a_{\Omega'}^{\dagger}} = 2 \pi \delta(\Omega - \Omega') \, , \qquad \comm*{b_{\omega}}{b_{\omega'}^{\dagger}} = 2 \pi \delta(\omega - \omega') 
\ee 
for the creation and annihilation operators associated with the modes inside and outside the shell, respectively. 
The initial state of the field is assumed to be the vacuum state $\ket{0}$ inside the shell defined by
\be 
\aOm \ket{0} = 0 \quad \forall \ \Omega \in(0, \Lambda) \, .
\ee 
The quantity relevant to Hawking radiation is the number of Hawking particles 
in the state $|0\rangle$ at large $r$.
From eqs.~\eqref{eq:bogoliubov} and~\eqref{eq:alphabeta}, we find
\be 
\ev{b_{\omega}^\dagger b_{\omega'}}{0} = \int_0^{\Lambda}\frac{d\Omega}{2\pi} \, \beta^*_{\omega \Omega} \, \beta_{\omega' \Omega} = \int_0^{\Lambda}\frac{d\Omega}{2\pi} \, \frac{8\Omega}{g_{\Omega}'} \, \widetilde{\varphi}_{\omega} (-p_\Omega) \, \widetilde{\varphi}_{\omega'}^{\,*} (-p_\Omega) \, .
\ee
Plugging in the solution~\eqref{eq:wavesoln} with the change of variable $q = -p_{\Omega}$, we arrive at
\be 
\label{eq:vev_mono0}
\begin{aligned}
    \ev{b_{\omega}^\dagger b_{\omega'}}{0} = \mathcal{N}_{\omega} \, \mathcal{N}_{\omega'} \int^{-\infty}_{0} &\frac{dq}{2\pi} \, \frac{2\,g(q)}{\bigl( 2\omega - g(q) \bigr) \bigl( 2\omega' - g(q) \bigr)} \times \\
    & \times \exp\left[ - 2i a (\omega - \omega') \int_0^q dk \, \frac{g(k)}{\bigl( 2\omega - g(k) \bigr) \bigl( 2\omega' - g(k) \bigr)} \right] \, .
\end{aligned}
\ee
By introducing a new variable 
\be 
\label{eq:F}
F(q)\equiv \int_0^{q}dk \, \frac{g(k)}{\bigl( 2\omega - g(k) \bigr) \bigl( 2\omega' - g(k) \bigr)} \, ,
\ee
with $F(-\infty) = \infty$ due to the fact that $g(p)$ is odd, monotonic, and that $|g(p)| \leq |p|$,
eq.~\eqref{eq:vev_mono0} simplifies to
\begin{align}
\label{eq:vev_mono}
\ev{b_{\omega}^\dagger b_{\omega'}}{0} 
&=
\mathcal{N}_{\omega} \, 
\mathcal{N}_{\omega'} \int^{\infty}_{0} \frac{dF}{\pi} \, 
e^{- 2 i a (\omega - \omega') F} 
= 
\mathcal{N}_{\omega} \, \mathcal{N}_{\omega'} \, 
\frac{1}{2 \pi i a (\omega - \omega' - i 0)} \, .
\end{align}
Interestingly, we observe that the modifications encoded in $g(p)$ are entirely hidden within the new variable $F$.
In other words, from the viewpoint of Hawking radiation, the effect of modifying the dispersion relation amounts to a mere relabeling of trans-Planckian modes. 
This explains why Hawking radiation remains intact despite such modifications.
Hawking radiation is thus insensitive to monotonic UV modifications to the dispersion relation, 
a result consistent with the findings of Ref.~\cite{Brout:1995wp}. 
For a more physical description of Hawking radiation as a transient phenomenon,
one should replace the single-frequency modes by wave packets~\cite{Hawking:1975vcx}. 
This enables us to discuss the detection of a localized Hawking particle.
Consequently, we will be able to examine the magnitude of radiation at different times.
An outgoing Hawking wave packet with central Killing frequency $\omega_0 \sim \mathcal{O}(1 / a)$ can be constructed as
\be 
\label{eq:packet}
\Psi_{(\omega_0,u_0)}(v, r) = 
\int\frac{d\omega}{2\pi} \, 
f_{\omega_0}(\omega) \, e^{i\omega u_0} \, \phi_\omega(v, r) \, ,
\ee
where $f_{\omega_0}(\omega)$ represents a narrow profile 
with its peak at $\omega_0$ and a width of $\Delta \omega \ll \omega_0$.
This profile function $f_{\omega_0}(\omega)$ specifies the relative amplitudes of the frequency eigenmodes $\phi_\omega$ in the linear superposition\,\footnote{
Commonly used profiles include the Gaussian profile 
as well as the step function profile $f_{\omega_0}(\omega) = \left[ \Theta(\omega - \omega_0 + \Delta \omega / 2) - \Theta(\omega - \omega_0 - \Delta \omega / 2) \right] / \sqrt{\Delta \omega}$,
which was originally adopted by Hawking in his seminal work~\cite{Hawking:1975vcx}.
}.
According to the spatial asymptotic behavior $\phi_\omega(v, r \to \infty) \propto e^{-i\omega u}$, 
the wave packet~\eqref{eq:packet} is approximately a function of 
the light-cone retarded time $u = v - 2r_*$ at large distances.
The factor $e^{i \omega u_0}$ shifts the center of the wave packet to $u = u_0$.
The creation operator corresponding to the Hawking quanta~\eqref{eq:packet} is
\be
b_{\Psi}^{\dagger} = \int\frac{d\omega}{2\pi} \, f_{\omega_0}(\omega) \, e^{i\omega u_0} \, b_{\omega}^{\dagger} \, ,
\ee
where $f_{\omega_0}(\omega)$ has been suitably normalized so that 
\be 
\comm*{b_\Psi}{b_{\Psi}^{\dagger}} = \int\frac{d\omega}{2\pi} \, \abs{f_{\omega_0}(\omega)}^2 = 1 \, .
\ee 
The number of Hawking particles with approximate frequency $\omega_0$ 
detected around the retarded time $u = u_0$ 
by an asymptotic observer is then given by the vacuum expectation value
\be 
\label{eq:HR}
\ev{b_{\Psi}^{\dagger} b_\Psi}{0} = \int\frac{d\omega}{2\pi} \int\frac{d\omega'}{2\pi} \, f_{\omega_0}(\omega) f_{\omega_0}^*(\omega') \, e^{i (\omega - \omega') u_0} \ev{b_{\omega}^{\dagger} b_{\omega'}}{0} \, .
\ee
Inserting eq.~\eqref{eq:vev_mono} into the above and then pulling out the slowly-varying factor in the integrand, which can be approximated by its value at $\omega_0$, we find 
\begin{align}
    \ev{b_{\Psi}^{\dagger} b_\Psi}{0} 
    &\approx
    \frac{1}{e^{4\pi a \omega_0 / g_{\omega_0}'} - 1} \, \int\frac{d\omega}{2\pi} \int\frac{d\omega'}{2\pi} \, f_{\omega_0}(\omega) f_{\omega_0}^*(\omega') \, e^{i (\omega - \omega') u_0} \frac{1}{i (\omega - \omega' - i 0^+)} 
    \nn \\
    &= 
    \frac{1}{e^{4\pi a \omega_0 / g_{\omega_0}'} - 1} \, \int\frac{d\omega}{2\pi} \int\frac{d\omega'}{2\pi} \, f_{\omega_0}(\omega) f_{\omega_0}^*(\omega') \int_{-\infty}^{u_0} du \, e^{i (\omega - \omega') u} 
    \nn \\
    &= 
    \frac{1}{e^{4\pi a \omega_0 / g_{\omega_0}'} - 1} \, \int_{-\infty}^{u_0} du \, \big\lvert \mathcal{F}\{ f_{\omega_0} \} (u) \big\rvert^2 \, , 
    \label{eq:VEV_mono}
\end{align}
where $\mathcal{F}\{f_{\omega_0}\}$ is the Fourier transform of $f_{\omega_0}$, 
and the integration variable $u$ can be identified with the retarded time.
The width of $\mathcal{F}\{f_{\omega_0}\}$ in the $u$-space is the inverse of the width of $f_{\omega_0}$ 
in the $\omega$-space (e.g. $\Delta u \sim 100a$ for $\Delta\omega \sim 0.01a^{-1}$). 
Therefore, the integral in eq.~\eqref{eq:VEV_mono} is approximately $0$ for $u_0 \ll - \Delta u$,
and approximately $1$ for $u_0 \gg \Delta u$.
Hence, there is a steady flux of Hawking quanta once $u_0 \gg 1/\Delta\omega$
(at least before the black hole becomes microscopic).
Note that,
since $g'_{\omega} \approx 1$ so long as $\omega \ll M_p$,
the factor $1/(e^{4\pi a\om/g'_{\om}} - 1)$ of Planck distribution in eq.~\eqref{eq:VEV_mono}
indicates a robust Hawking temperature $(4 \pi a)^{-1}$ 
regardless of the time-dependent magnitude of Hawking radiation.
For the standard dispersion relation, a wave packet with a well-defined frequency $\omega \sim \mathcal{O}(1 / a)$ 
(and a small uncertainty $\Delta \omega$) has a large $\Delta u \gtrsim \mathcal{O}(a)$.
Due to the exponential relation $U \simeq -2 a e^{-u / 2a}$ between the retarded time $u$ for distant observers and the Minkowski retarded time $U$ inside the shell, 
this wave packet is highly blue-shifted in the past and is confined within a tiny region $\Delta U$.
This implies a large uncertainty $\Delta\Omega$ in its frequency $\Omega$, resulting in a mixture of positive and negative-frequency modes, and thus the appearance of Hawking radiation.
As explained earlier, 
the robustness of Hawking radiation demonstrated in eq.~\eqref{eq:VEV_mono} 
stems from the fact that the UV-modification of monotonic dispersion relations 
can be interpreted as a mere relabeling of states, 
without altering the composition of positive and negative-$\Omega$ modes within the wave packet.
On the other hand,
we also note the possibility that,
if the range of the $F$-integration is finite,
Hawking radiation can still be modified,
although the Planck distribution at Hawking temperature remains the same.
The range of the $F$-integration in eq.~\eqref{eq:vev_mono} is in general $\bigl( 0, F(-\infty) \bigr)$,
and $F(-\infty)$ can be finite if either $g(p) \rightarrow \infty$ or $g(p) \rightarrow 0$
sufficiently fast in the limit $p \rightarrow \infty$.
The former possibility corresponds to a \emph{superluminal} dispersion relation outside the horizon,
whereas the latter corresponds to a non-monotonic dispersion relation.
We shall study the latter case in the next section,
and leave the first possibility for future study.

\section{Non-Monotonic Dispersion}
\label{sec:nonmonotonic}

\begin{figure}[ht]
\centering
\includegraphics[scale=1.0]{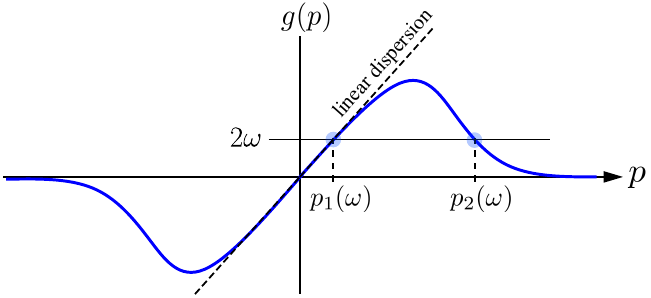}
\caption{An illustration of the non-monotonic dispersion curve under consideration. 
There exists a maximum $\Lambda = \mathrm{max} \{ g(p) / 2 \}$ in energy 
but not necessarily in momentum. 
For $0 < \omega < \Lambda$, 
the dispersion relation $2 \omega = g(p)$ admits two solutions $p_1(\omega)$ and $p_2 (\omega)$.
The trans-Planckian mode $p_2 (\omega)$ has negative group velocity since $g' (p \gtrsim M_p) < 0$.}
\label{fig:k}
\end{figure}

In this section,
we study Hawking radiation for non-monotonic dispersion relations
with a single maximum $\Lam \equiv \mbox{max}\{g(p)/2\}$
and $g(p) \rightarrow 0$ in the limit $p\rightarrow\infty$
(see Fig.~\ref{fig:k}).
Such UV dispersions add complexity to the analysis 
by introduction of an additional ingoing mode.
As depicted in Fig.~\ref{fig:k}, 
for any given $0 < \omega < \Lambda$, 
there are two solutions $p_i (\omega)$ ($i = 1, 2$) to the dispersion relation:
\be 
\label{eq:p1p2}
2 \omega = g(p_1) = g(p_2) \, , \qquad \text{where} \ \ p_1 (\omega) < p_{\Lam} < p_2(\omega) \, ,
\ee 
with $p_{\Lam}$ being the momentum at the maximum frequency, i.e. 
\be
2\Lambda = g(p_{\Lambda}) \, .
\ee
Realistic dispersion relations can be even more complicated, 
possibly involving more than two solutions. 
For instance, 
in analogue black-hole systems, 
the dispersion relation for excitations may feature a roton minimum~\cite{Jacobson:1991gr}, 
which was recently shown~\cite{Ribeiro:2022gln} to have a significant impact on the radiation spectrum.

If we trace a wave packet back in time, 
the negative group velocity in the UV regime of Fig.~\ref{fig:k} 
means that the wave packet bounces back from the horizon to great distances (in the infinite past).
At first sight,
this would seem to lead to the suppression of Hawking radiation,
as the relation between the affine parameter $u$ in the infinite future
and the affine parameter in the infinite past is not of an exponential form.
However,
it turns out that the non-monotonic dispersion relation 
also allows for tunneling across the horizon.
Taking the tunneling effect into account,
we find that Hawking radiation is either unchanged or significantly modified,
depending on whether the dispersion curve decays faster than $\mathcal{O}(1/p)$
in the large momentum limit ($p \rightarrow \infty$).

\subsection{Time-dependence of Hawking radiation}
\label{sec:HRoff}

The following calculation of Hawking radiation
is a slight modification of Sec.~\ref{sec:HR_mono},
except that the dispersion relation is non-monotonic here.
Outside the shell, 
the decomposition~\eqref{eq:expand_b} and the quantization of the field $\phi(v > 0, r)$ remain valid\,\footnote{
To adopt the results of Sec.~\ref{sec:HR_mono},
we also need to replace $p_{\om}$ by $p_1(\om)$.
For instance,
$g'_{\omega}$ in eq.~\eqref{eq:g_prime} should be replaced by
$g'_1(\omega) \equiv g'\bigl( p_1(\omega) \bigr)$.
}.
Inside the shell,
since there are two solutions of $p$~\eqref{eq:p1p2} with the same frequency $\Omega$
(denoted by $p_1(\Omega)$ and $p_2(\Omega)$),
the basis modes and the creation-annihilation operators  
are divided into two sets as
$\bigl\{ \chi_{\Omega}^1 (v, r), \chi_{\Omega}^2 (v, r) \bigr\}$ and 
$\bigl\{ a_{\Omega}^1, a_{\Omega}^{1 \, \dagger}, a_{\Omega}^2, a_{\Omega}^{2 \, \dagger}\bigr\}$,
respectively.
The field inside the shell can then be expanded as 
\be 
\phi(v < 0, r) = \int_0^{\Lambda}\frac{d\Omega}{2\pi} \, \sum_{i = 1}^2 \left[ a_{\Omega}^i \, \chi_{\Omega}^i (v, r) + a_{\Omega}^{i \, \dagger} \, \chi_{\Omega}^{i \, *} (v, r) \right] \, ,
\ee 
where 
\be 
\chi_{\Omega}^i (v, r) = \frac{1}{\sqrt{2\Omega \, \abs{g_i' (\Omega)}}} \, e^{-i \Omega v \, + \, i p_i (\Omega) r} \, .
\ee 
Here, we have used the shorthand notation
\be 
g_i'(\Omega) \equiv \left.\frac{dg(p)}{dp}\right|_{p \, = \, p_i(\Omega)} \, .
\ee 
The creation and annihilation operators obey the commutation relation $\comm*{a_{\Omega}^i}{a_{\Omega'}^{j \, \dagger}} = 2 \pi \, \delta_{ij} \delta(\Omega - \Omega')$, 
and the Minkowski vacuum is defined by
\be 
a_{\Omega}^i \ket{0} = 0 \quad \forall \ \Omega \in(0, \Lambda) \ \text{and} \ i \in \{ 1, 2 \} \, .
\ee 
The continuity of $\phi(v, r)$ across the null shell located at $v = 0$ implies the linear relation
\be 
b_\omega = \int_0^\Lambda \frac{d\Omega}{2\pi} \sum_{i = 1}^2 \left[ \alpha_{\omega\Omega}^i \, a_{\Omega}^i + \beta_{\omega\Omega}^i \, a_{\Omega}^{i \, \dagger} \right] \, ,
\ee 
where the Bogoliubov coefficients are found to be
\be 
\alpha_{\omega\Omega}^i = 
2 \sqrt{\frac{2\Omega}{\abs{g_i' (\Omega)}}} \, 
\widetilde{\varphi}_\omega^{\,*} \bigl( p_i (\Omega) \bigr) \, , \qquad \beta_{\omega\Omega}^i 
= 
- 2 \sqrt{\frac{2\Omega}{\abs{g_i' (\Omega)}}} \, 
\widetilde{\varphi}_\omega^{\,*} \bigl( -p_i (\Omega) \bigr) \, .
\ee 
This is a straightforward generalization of eq.~\eqref{eq:alphabeta}.
With the change of variable $q = - p_i (\Omega)$, 
we find
\begin{align}
\ev{b_{\omega}^\dagger b_{\omega'}}{0} 
&= 
\int_0^{\Lam} \frac{d \Omega}{2 \pi} \, 
\sum_{i = 1}^2 \beta_{\omega\Omega}^{i \, *} \, \beta_{\omega'\Omega}^i 
\nn \\
&=
\left[ \int_0^{- p_{\Lambda}} + \int_{- p_{\Lambda}}^{-\infty} \right]
\frac{dq}{2 \pi} \,2 g(q) \, 
\widetilde{\varphi}_\omega (q) \, \widetilde{\varphi}_{\omega'}^{ \, *} (q) 
\nn \\
&= \mathcal{N}_{\omega} \, \mathcal{N}_{\omega'} \int_0^{F(-\infty)} \frac{dF}{ \pi} \, e^{-2 i a (\omega - \omega') F} \, ,
\label{eq:vev_nonmono}
\end{align}
where $F(q)$ is the change of variable introduced in eq.~\eqref{eq:F}. 
Formally, the expression~\eqref{eq:vev_nonmono} is identical to 
eq.~\eqref{eq:vev_mono0} for monotonic dispersion relations.
However,
the upper bound $F(-\infty)$ of the integration over $F$ could be different.
While $F(-\infty)$ is always infinite for subluminal monotonic dispersion relations,
it can be finite for non-monotonic dispersion relations.
The expectation value of the number of Hawking particles detected around $u = u_0$
can be evaluated as
\begin{align}
\ev{b_{\Psi}^{\dagger} b_\Psi}{0} &\approx \frac{2 \pi a}{e^{4\pi a \omega_0 / g_1'(\omega_0)} - 1} \, \int\frac{d\omega}{2\pi} \int\frac{d\omega'}{2\pi} \, f_{\omega_0}(\omega) f_{\omega_0}^*(\omega') \, e^{i (\omega - \omega') u_0} \int_0^{F(-\infty)} \frac{dF}{\pi} \, e^{-2 i a (\omega - \omega') F} \nn \\
&= \frac{1}{e^{4\pi a \omega_0 / g_1'(\omega_0)} - 1} \, \int\frac{d\omega}{2\pi} \int\frac{d\omega'}{2\pi} \, f_{\omega_0}(\omega) f_{\omega_0}^*(\omega') \int_{u_0 - 2 a F(-\infty)}^{u_0} du \, e^{i (\omega - \omega') u} \nn \\
&\approx \frac{1}{e^{4\pi a \omega_0 / g_1'(\omega_0)} - 1} \, \int_{u_0 - u_{\Delta}}^{u_0} du \, \big\lvert \mathcal{F}\{ f_{\omega_0} \} (u) \big\rvert^2 \, , 
\label{eq:VEV_nonmono}
\end{align}
where 
\begin{align}
u_{\Delta} = 2a F_0(-\infty) &\equiv 
2a \int_0^{-\infty} dp \, \frac{g(p)}{\left[ 2\omega_0 - g(p) \right]^2} \nn \\
&\approx 2a \int_0^{-M_p} dp \, \frac{p}{(2\omega_0 - p)^2} + 2a \int_{-M_p}^{-\infty} dp \, \frac{g(p)}{\left[ 2\omega_0 - g(p) \right]^2} \nn \\
&\approx 2a\ln \left( \frac{M_p}{2\omega_0} \right) + 2a \int_{-M_p}^{-\infty} dp \, \frac{g(p)}{\left[ 2\omega_0 - g(p) \right]^2} \, .
\label{eq:turnoff_t}
\end{align}
The first term in eq.~\eqref{eq:turnoff_t} 
is roughly the scrambling time of a black hole for $\omega_0 \sim \mathcal{O}(1 / a)$.
The second term depends on the dispersion relation.
Its physical interpretation will be discussed in Sec.~\ref{sec:tunneling}. 
By assumption,
the center of the function $\mathcal{F}\{f_{\omega_0}\}(u)$ is at the origin $u = 0$ in the $u$-space
(see eq.~\eqref{eq:packet}).
Therefore, 
for Hawking particles at late times satisfying $u_0 \gg u_{\Delta}$,
the integral in eq.~\eqref{eq:VEV_nonmono} becomes negligibly small, 
indicating that Hawking radiation is turned off for a distant observer 
after a certain retarded time $u \sim u_{\Delta}$\,\footnote{
More precisely,
for a given wave packet with a width $\Delta u$,
the integral in eq.~\eqref{eq:VEV_nonmono}
becomes negligible when $u_0 - u_{\Delta} \gg \Delta u$.
For the dominant frequency $\omega_0 \sim 1/a$,
the wavelength is $\sim a$,
and the wave packet has a width $\Delta u \gg a$
(e.g. $\Delta u \sim 100 a$).
In the sense of the $(a M_p)^{-1}$-expansion,
$\Delta u$ is of order $\mathcal{O}(a)$,
whereas $u_{\Delta} \gg \mathcal{O}(a)$, as we will see shortly.
It is thus not important to keep $\Delta u$ here.
}.  
The time scale for turning off Hawking radiation
is closely linked to the rate at which $g(p)$ approaches zero in the limit $p \rightarrow \infty$. 
To understand this connection,
we consider the toy model
\be 
g(p) = 
\begin{cases}
    M_p^{n + 1} \, p^{-n} & \text{for} \ p > M_p \, , \\
    p & \text{for} \ 0 \leq p \leq M_p \, , \\
    - g \bigl( \abs{p} \bigr) & \text{for} \ p < 0 \, ,
\end{cases}
\ee 
which gives 
\be 
u_{\Delta} 
\approx 2a \ln \left(\frac{M_p}{2\omega_0}\right) + \frac{\pi}{n^2 \sin(\pi/n)} \, 2 a \left(\frac{M_p}{2\omega_0}\right)^{1 \, + \, 1/n} \, .
\ee 
The critical value is $n=1$,
for which the second term becomes infinite, 
and Hawking radiation~\eqref{eq:VEV_nonmono} persists,
similar to the case of subluminal monotonic dispersion relations.
However, for $n > 1$ and for Hawking radiation with the dominant frequency $\omega_0 \sim 1/a$,
the second term is of order $\mathcal{O}\bigl( a ( M_p \, a )^{1 \, + \, 1 / n} \bigr)$, 
which is much shorter compared to $\mathcal{O}\bigl( M_p^2 \, a^3 \bigr)$ 
for a large black hole with $M_p \, a \gg 1$. 
As a result, 
Hawking radiation is turned off well before the Page time $\mathcal{O}\bigl( M_p^2 \, a^3 \bigr)$~\cite{Page:1993wv} for $n > 1$.
The example above demonstrates how the duration of Hawking radiation can be adjusted
by tuning the dispersion curve $g(p)$ at large $p$. 
By considering a dispersion curve that exhibits a much more rapid decay, 
such as $g(p) \propto p \, e^{-p^{2n}}$,
Hawking radiation is terminated around the scrambling time, 
resembling the effect of a UV cutoff~\cite{Ho:2022gpg}.
Another example is the Corley-Jacobson dispersion relation considered in Ref.~\cite{Corley:1996ar}:
\be 
g(p) = p \sqrt{1 - \left( p / M_p \right)^{2n}} \, .
\ee 
In this case, the integral in eq.~\eqref{eq:VEV_nonmono} takes the form
\be 
\int_{u_0 - 2 a F_0 ( - M_p)}^{u_0} du \, \big\lvert \mathcal{F}\{ f_{\omega_0} \} (u) \big\rvert^2 \, ,
\ee 
where the lower bound implies that 
Hawking radiation comes to a stop after $u \sim 2 a F_0 (- M_p)$, 
which is also of the order of the scrambling time\,\footnote{This observation, 
that the presence of a momentum cutoff in the dispersion relation 
leads to the termination of Hawking radiation, 
was also pointed out in Ref.~\cite{Brout:1998ei}.
It can be viewed as a special case in our formulation
by setting $g(p \geq p_{\mathrm{cut}}) = 0$ for a given cutoff momentum $p_{\mathrm{cut}}$.
}.

\subsection{Bouncing off horizon}
\label{sec:bounce}

In the above,
we found that Hawking radiation is turned off after $u \sim u_{\Delta}$ 
for non-monotonic dispersion relations with $g(p)$ decaying faster than $\mathcal{O}(1/p)$ as $p \rightarrow \infty$.
We investigate the underlying physics from the perspective of a wave packet
in this and the next subsections.
We will again trace a Hawking wave packet backward in time, 
and observe the following.
Initially, the wave packet approaches the horizon, 
and then it branches off into two parts. 
A part of it is reflected back towards large $r$, 
while the other tunnels through the horizon and continues to propagate towards the singularity at $r = 0$.
We examine the branch that bounces off the horizon in this subsection.
In the regime of the phase space where the WKB approximation holds, the trajectory $\bigl( x(v), p(v) \bigr)$ followed by the wave packet aligns with the classical particle trajectory, which can be determined from eqs.~\eqref{eq:hamiltonian}--\eqref{eq:hamilton2}. 
In particular, making use of eq.~\eqref{eq:disprel_out},
we have 
\begin{align}
    \label{eq:dpdv}
    \frac{dp}{dv} &= - \frac{1}{2a} \, \frac{\left[ g(p) - 2 \omega \right]^2}{g(p)} \, , \\
    \frac{dx}{dv} &= \frac{\omega}{g(p)} \, g'(p) \, .
    \label{eq:dxdv}
\end{align}

Consider the backward evolution of a particle in the state 
$(x, p) \simeq (\infty, \, p_1(\omega) > 0)$ from $v = \infty$. 
According to eq.~\eqref{eq:dpdv}, 
the momentum of the particle would increase monotonically as $v$ decreases, 
resulting in a blueshift that continues backward in time 
until eventually $p \to p_2(\omega)$ in the infinite past. 
In terms of the particle's position $x$, 
when $p \lesssim M_p$, 
the particle follows a trajectory similar to that dictated by the standard dispersion relation. 
When the momentum becomes Planckian 
and the dispersion curve starts to drop (i.e. $g'(p \gtrsim M_p) < 0$), 
which occurs when the particle is approximately a Planck length outside the horizon, 
the particle's velocity is reversed according to eq.~\eqref{eq:dxdv}. 
This causes the particle to be ``bounced off'' from the horizon 
and move away towards large distances, 
as shown in Fig.~\ref{fig:l} below. 
A straightforward calculation using eq.~\eqref{eq:dxdv0} 
with a negative $g'(p \gtrsim M_p)$ reveals that 
the particle will bounce back and cross the null shell 
at a distance $x \gtrsim \mathcal{O}(a)$
if the retarded time $u_0$ at which the particle reaches future infinity 
is greater than the scrambling time, i.e. $u_0 \gtrsim \mathcal{O} \bigl( a \log(M_p \, a) \bigr)$.
The continuity condition of the field is imposed on the matter shell
to determine whether this wave packet contains negative-frequency modes 
in the Minkowski space inside the shell.
Since the ingoing wave packet 
has a finite extent $\Delta x \sim \mathcal{O}(a)$ on the shell,
it has a small spread $\Delta p \sim \mathcal{O}(1/a)$ 
in its momentum $p \gtrsim M_p$ without any negative-$p$ components in the mix.
Hence,
for this bouncing branch,
a positive-$\omega$ outgoing wave packet is originated from
an ingoing wave packet with purely positive frequency $\Omega$ inside the shell.
The bouncing branch of the wave packet does not contribute to Hawking radiation. 
\begin{figure}[ht]
\centering
\includegraphics[scale=1.0]{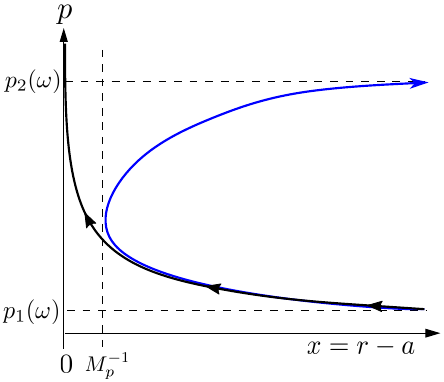}
\caption{A sketch of the characteristic phase space trajectories outside the horizon backward in time.
For subluminal monotonic dispersion relations, the trajectory (black curve) converges toward some fixed radius. 
For non-monotonic dispersion relations, the particle turns around at a Planckian distance from the horizon and then travels towards large distances, eventually approaching the momentum $p_2 (\omega)$.}
\label{fig:l}
\end{figure}
Contrary to Ref.~\cite{Brout:1995wp}, 
where an analysis of eq.~\eqref{eq:wave_eqn_out} 
in the near-horizon region $x \equiv r - a \ll a$ is sufficient, 
here we have to expand our discussion to $x \gtrsim \mathcal{O}(a)$.
Therefore, 
only by solving the wave equation across the entire space of $r > 0$ 
(as we did above)
can we properly account for Hawking radiation at late times. 
Naively, 
one might expect no Hawking radiation at all 
due to this bounce-off behavior for late-time Hawking modes. 
However, 
as hinted above, 
there is more to the story, 
on which we will elaborate in the following subsection.

\subsection{Tunneling across horizon}
\label{sec:tunneling}

\begin{figure}[ht]
\centering
\includegraphics[width=0.9\textwidth]{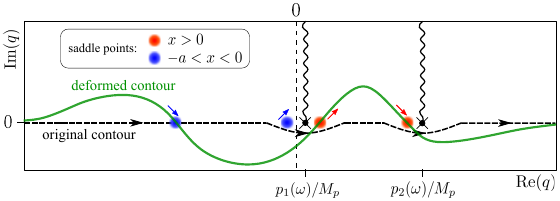}
\caption{A sketch of the integration contours in the complex-$q$ plane for eq.~\eqref{eq:inv_FT}. 
The saddle points in the regions $\abs{x} \gg M_p^{-1}$ are marked, with the arrows indicating the directions of steepest descent. 
The original contour (dashed black curve) corresponds to the $i0^+$-prescription that defines the Hawking modes satisfying the appropriate boundary conditions. 
It can be continuously deformed into the green curve while preserving its topology with respect to the singularities and branch cuts. 
For positive $x$, the deformed contour passes through both positive-momentum saddle points along their respective steepest-descent directions. 
Beyond $x \ll - M_p^{-1}$, the integral receives a contribution only from the saddle point with a large negative momentum.}
\label{fig:saddle}
\end{figure}

\begin{figure}[ht]
\centering
\includegraphics[scale=1.0]{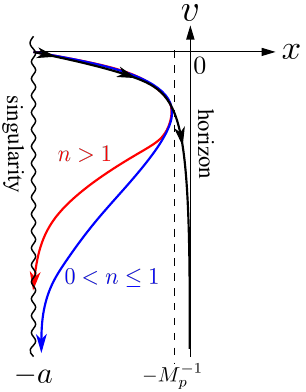}
\caption{A sketch of the classical trajectories inside the horizon. The arrows indicate backward evolution. The black curve represents the trajectory associated with the standard dispersion relation. For non-monotonic dispersion profiles that decay as $g(p) \sim M_p^{n + 1} \, p^{-n}$ for large $p$, the corresponding trajectories are depicted. They exhibit a turning point near $x \sim - M_p^{-1}$ and a segment of superluminal propagation towards the singularity. The red curve reaches the singularity in a finite amount of time, while the blue curve approaches it asymptotically.}
\label{fig:traj_inside}
\end{figure}

When the wave packet is traced back to the turning point of the bouncing trajectory, 
its spatial profile is highly compressed.
Therefore, 
the large uncertainty $\Delta p$ in momentum 
has to be taken into consideration, 
rendering the particle description of the wave packet inadequate. 
To analyze the situation properly, 
we turn to the wave solutions in the position space:
\be 
\label{eq:inv_FT}
\phi_{\omega}(v, x) = 
\mathcal{N}_{\omega} \, e^{- i \om v}
\int_{-\infty}^{\infty} \frac{dq}{2 \pi} \, 
\frac{M_p \, e^{i M_p \, x q}}{2 \omega - g(M_p \, q - i 0^+)} \, \exp \left[ 2 i a \omega \int_0^q \frac{M_p \, ds}{2 \omega - g(M_p \, s - i 0^+)} \right] \, ,
\ee 
where $q \equiv p / M_p$ and $s$ are dimensionless variables. 
In the region where $\abs{M_p \, x} \gg 1$, 
the integral can be effectively approximated by the contributions from the saddle points $q_s = p_s (x) / M_p$, which coincide with the solutions to the dispersion relation~\eqref{eq:disprel_out}. 
In the limit $x \to \infty$, 
two positive-momentum saddle points (see Fig.~\ref{fig:saddle})
give rise to the outgoing Hawking mode $p_1 (\omega)$ at future infinity and the ingoing wave $p_2 (\omega)$ at past infinity, respectively.
This corresponds to the portion of the wave that evolves along the bouncing branch, 
which we have argued not to contribute to Hawking radiation in Sec.~\ref{sec:bounce}.
Interestingly, 
inside the horizon where $-a < x \ll -M_p^{-1}$, 
there exists an additional saddle point with negative momentum (see Fig.~\ref{fig:saddle}).
In the point particle description, 
this saddle point corresponds to a superluminal trajectory inside the horizon, as illustrated in Fig.~\ref{fig:traj_inside}.
As a result, 
not all of the wave is bounced away from the black hole 
for non-monotonic dispersion relations; 
instead, a portion of it ``tunnels'' through the horizon. This is reminiscent of the interpretation of
Hawking radiation as the tunneling of point particles across
the horizon along a classically forbidden trajectory \cite{Parikh:1999mf}. It
occurs through a different mechanism that involves the
backreaction of radiation, while the tunneling here arises
due to superluminal dispersive propagation.
This phenomenon, where a positive-$\omega$ wave packet is partially converted into a negative-momentum wave packet, is known as \emph{mode conversion}~\cite{Corley:1996ar}. 
It takes place in the vicinity of the turning point and is the primary mechanism responsible for the creation of Hawking radiation in our case.
Similar effects have been found in previous works~\cite{Brout:1995wp, Corley:1996ar, Corley:1996nw, Corley:1997pr, Himemoto:1999kd, Unruh:2004zk} that considered both monotonic and non-monotonic dispersion relations in the freely falling frame.
However, there are notable differences.
In those works, the converted wave packet with negative free-fall frequency
has a negative group velocity and is supported outside the horizon, 
leading to the conclusion that a late-time, positive-$\omega$ wave packet 
is originated from a pair of ingoing wave packets with opposite signs of trans-Planckian momenta at large $r$.
In contrast, in our current work, the negative-momentum component resides inside the horizon and approaches the singularity as it propagates backwards in time. Consequently, non-monotonic dispersion relations in the Eddington-Finkelstein frame present an intriguing system in their own right, as the mode conversion process enables the wave to tunnel between two branches of causally disconnected characteristic curves.
Depending on whether the negative-momentum wave packet reaches the singularity within a finite time, it could have significant implications for the late-time behavior of Hawking radiation. 
In the region $\abs{M_p \, x} \gg 1$ inside the horizon, this wave packet is well-localized along the classical trajectory described by eqs.~\eqref{eq:disprel_out}, \eqref{eq:dpdv}, and \eqref{eq:dxdv}. 
During the backward time evolution, eq.~\eqref{eq:dpdv} indicates that the central momentum of the wave packet evolves from $p \simeq -M_p$ to $p = - \infty$ within the time interval
\be 
\label{eq:v_singularity}
\abs{\Delta v} = 2 a \int_{-M_p}^{-\infty} dp \, \frac{g(p)}{\left[ g(p) - 2 \omega \right]^2} = 2 a \int_{M_p}^{\infty} dp \, \frac{g(p)}{\left[ g(p) + 2 \omega \right]^2} \, .
\ee 
In terms of how $x$ and $p$ are related along the characteristic phase space curve~\eqref{eq:disprel_out}, $\abs{\Delta v}$ in fact also represents the time it takes for the peak of the wave packet to travel from the turning point $x \sim - M_p^{-1}$ inside the horizon to the singularity. 
Notice that the expression~\eqref{eq:v_singularity} is precisely the second term in eq.~\eqref{eq:turnoff_t} that plays a crucial role in determining whether Hawking radiation will be largely suppressed after a retarded time $u \sim u_{\Delta}$.
The duration $\abs{\Delta v}$~\eqref{eq:v_singularity} depends heavily on the asymptotic behavior of the dispersion curve $g(p)$. 
Suppose that for very large values of $p \gg M_p$, the non-monotonic dispersion curve decays as $g(p) \sim M_p^{n + 1} \, p^{-n}$ with $n > 0$.
Since $g(p) \ll 2 \omega$ in this regime, the integrand is approximately linear in $g(p)$.
It is then evident from power counting that $\abs{\Delta v}$ is finite if $g(p)$ decreases faster than $\mathcal{O}(1/p)$ as $p$ approaches infinity (see Fig.~\ref{fig:traj_inside}).
Conversely, if $g(p)$ does not decrease sufficiently fast (i.e. $n \leq 1$), the duration $\abs{\Delta v}$ becomes infinite, indicating that the negative-momentum wave packet would ultimately reach the matter shell, 
resulting in the production of Hawking radiation.
\begin{figure}[ht]
\centering
\includegraphics[scale=1.1]{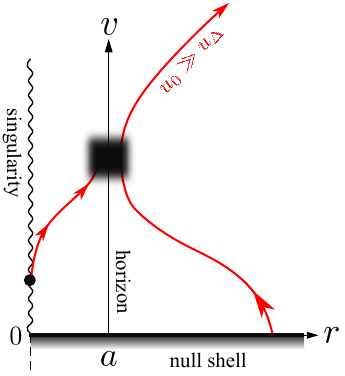}
\caption{A sketch of the trajectory of a wave packet that reaches asymptotic infinity at a sufficiently late retarded time $u_0 \gg u_{\Delta}$. 
The shaded region represents the breakdown of the particle description. 
In the case of non-monotonic dispersion curves that decay faster than $\mathcal{O}(1/p)$ for large values of $p$, 
the negative-momentum component of this wave packet, 
which tunnels across the horizon, 
is originated from the singularity.}
\label{fig:tunnel}
\end{figure}
In the case where $\abs{\Delta v}$ is finite, 
for a Hawking particle centered at a late time $u_0 \gg u_{\Delta}$, 
a portion of its wave packet is originated from the singularity at $r = 0$ (see Fig.~\ref{fig:tunnel}). 
In our calculations in Sec.~\ref{sec:HRoff}, 
the negative-momentum wave packet from the singularity is ignored,
as it does not involve any negative-frequency mode in the Minkowski space inside the shell.
In a consistent quantum theory of gravity,
the singularity is resolved, 
and the quantum state at the singularity would completely determine the expectation value 
$\langle 0 | b_{\Psi}^{\dagger} b_{\Psi} | 0 \rangle$ of the number of Hawking particles.
The tunneling effect across the horizon could potentially serve as a physical channel 
for extracting information about the collapsed matter. 
In any case,
Hawking radiation from the \emph{Unruh vacuum} is terminated after $u \sim u_{\Delta}$.

\section{Conclusion and Discussion}
\label{sec:conclude}

In this work, with an aim to explore the relationship between Hawking radiation 
and dispersive effects in the UV regime, 
we followed the formulation presented in Ref.~\cite{Brout:1995wp} 
and investigated the time dependence of Hawking radiation for a massless real scalar field. 
We began by examining the case of monotonic dispersion relations 
that only allow for subluminal propagation outside the black hole horizon. 
Interestingly, 
we discovered that the modifications introduced by these dispersion relations 
on Hawking quanta can be understood as a mere relabeling of the momentum
$q \rightarrow F(q)$~\eqref{eq:F}.
Consequently, 
the properties of Hawking radiation, including its temperature and strength in time, 
are robust against such monotonic UV dispersions.
We then turned to non-monotonic dispersion relations, 
and observed that the mixture of positive and negative-momentum components
in a Hawking wave packet has an origin different
from the conventional Hawking radiation. 
Specifically, 
the positive-momentum component arises from an ingoing trans-Planckian wave 
that bounces off the horizon, 
while the negative-momentum component is originated from the interior of the black hole
via a tunneling process across the horizon.
Further analysis revealed that 
the Hawking radiation for non-monotonic dispersion relations can be significantly altered 
if the dispersion curve $g(p)$ approaches zero faster than $\mathcal{O}(1/p)$ as $p$ approaches infinity.
In such cases, 
Hawking radiation becomes dependent on the quantum state at the singularity $r=0$ after a specific retarded time $u \sim u_{\Delta}$~\eqref{eq:turnoff_t}. 
This time $u_{\Delta}$ is the sum of the scrambling time
and the duration it takes a wave packet
to propagate from the singularity to the horizon. 
Its order of magnitude ranges from the scrambling time $\mathcal{O} \bigl( a \log(M_p \, a) \bigr)$ 
up to the Page time $\mathcal{O} \bigl( M_p^2 \, a^3 \bigr)$.
If there is no outgoing mode emerging from the singularity, 
Hawking radiation simply stops after $u \sim u_{\Delta}$.
While there are uncertainties regarding Hawking radiation due to our limited knowledge about the singularity, 
we can safely conclude that Hawking radiation will undergo significant modifications 
as long as the quantum state at the singularity differs from the Minkowski vacuum state.
Contrary to the prevailing notion in the literature 
that Hawking radiation is robust against UV-modifications of the dispersion relation, 
our findings demonstrate that Hawking radiation is actually only insensitive to certain types of UV-modifications, 
while it can be highly sensitive to other types, 
to the extent that the radiation can be completely suppressed after the scrambling time.
This conclusion aligns with recent studies~\cite{Ho:2020cbf, Ho:2021sbi, Ho:2022gpg}
indicating that Hawking radiation can be significantly modified beyond the scrambling time due to other UV properties of the effective field theory,
such as higher-derivative interactions with the background.
It is crucial that we have analyzed the time dependence of Hawking radiation.
We have observed that 
even when the magnitude of Hawking radiation dramatically changes over time,
the Hawking temperature remains the same.
This is perhaps part of the reason why the substantial modifications to Hawking radiation discussed in this work may have eluded previous investigations. 
The focus on the time dependence of the magnitude of Hawking radiation, 
in addition to its temperature, 
has allowed us to uncover this novel aspect.

\section*{Acknowledgement}

We thank Yosuke Imamura, Henry Liao, Naritaka Oshita, and Yuki Yokokura for valuable discussions. 
E.T.A. is supported by the Foundation for the Advancement of Theoretical Physics and Mathematics ``BASIS'' grant, by RFBR grant 19-02-00815 and joint RFBR-MOST grant 21-52-52004 MNT\underline{\hspace{0.5em}}a, and by Russian Ministry of education
and science.
T.L.C., P.M.H., W.H.S., and C.T.W. are supported in part 
by the Ministry of Science and Technology, R.O.C.
(MOST 110-2112-M-002 -016-MY3),
and by National Taiwan University. 
H.K. thanks Prof. Shin-Nan Yang and his family
for their kind support through the Chin-Yu chair professorship.
H.K. is partially supported by Japan Society of Promotion of Science (JSPS),
Grants-in-Aid for Scientific Research (KAKENHI)
Grants No.\ 20K03970 and 18H03708,
by the Ministry of Science and Technology, R.O.C. (MOST 111-2811-M-002-016),
and by National Taiwan University.

\appendix

\section{The Bogoliubov Transformation}
\label{app:bogoliubov}

In this appendix, we derive the explicit forms of the Bogoliubov coefficients defined in eq.~\eqref{eq:bogoliubov}. 
Since the Minkowski mode $\chiOm(v = 0, r)$ represents a Fourier mode 
with a given momentum $p_{\Omega}$, 
the calculation is analogous to performing a Fourier transform of $\varphi_\omega(r)$. 
We write 
\begin{align}
    \varphi_{\omega}(r) &= \int_{-\infty}^{\infty} \frac{d p_\Omega}{2\pi} \, \widetilde{\varphi}_\omega(p_\Omega) \, e^{ip_\Omega r} \nn \\
    &= \int_{-\Lambda}^{\Lambda}\frac{d\Omega}{2\pi} \, \frac{dp_\Omega}{d\Omega} \, \widetilde{\varphi}_\omega(p_\Omega) \, e^{ip_\Omega r} \nn \\
    &= \int_{0}^{\Lambda} \frac{d\Omega}{2\pi} \, \frac{2}{g_{\Omega}'} \left[ \widetilde{\varphi}_\omega(p_\Omega) \, e^{ip_\Omega r}+\widetilde{\varphi}_\omega(-p_\Omega) \, e^{-ip_\Omega r} \right] \, .
\end{align}
By substituting this into the mode expansion~\eqref{eq:expand_b} and matching it with eq.~\eqref{eq:expand_a} at $v = 0$, we obtain 
\be 
\label{eq:a_fourier}
\aOm = \int^{\Lambda}_0 \frac{d\omega}{2\pi} \, 2\sqrt{\frac{2\Omega}{g^\prime_\Omega}} \left[ b_\omega \, \widetilde{\varphi}_\omega(p_\Omega) + b_\omega^\dagger \, \widetilde{\varphi}_\omega^{\,*}(-p_\Omega) \right] \, .
\ee 
On the other hand, 
the commutation relation constrains the Bogoliubov coefficients to satisfy the normalization condition 
\be 
\label{eq:bogo_condition}
2\pi \delta(\omega - \omega') = \comm*{b_{\omega}}{b_{\omega'}^{\dagger}} = \int_0^\Lambda\frac{d\Omega}{2\pi} \left( \alpha_{\omega\Omega} \, \alpha_{\omega' \Omega}^*-\beta_{\omega\Omega} \, \beta_{\omega' \Omega}^* \right) \, .
\ee 
It then follows from eqs.~\eqref{eq:bogoliubov} and~\eqref{eq:bogo_condition} that the inverse Bogoliubov transformation is
\be 
\aOm = \int_0^{\Lambda}\frac{d\omega}{2\pi} \left( \alpha_{\omega\Omega}^* \, b_{\omega} - \beta_{\omega\Omega} \, b_{\omega}^\dagger \right) \, .
\ee 
Comparing the equation above with eq.~\eqref{eq:a_fourier}, we arrive at eq.~\eqref{eq:alphabeta}:
\be 
\alpha_{\omega \Omega} = 2\sqrt{\frac{2\Omega}{g_{\Omega}'}} \, \widetilde{\varphi}_{\omega}^{\,*}(p_\Omega) \, ,
\qquad 
\beta_{\omega \Omega} =-2\sqrt{\frac{2\Omega}{g_{\Omega}'}} \, \widetilde{\varphi}_{\omega}^{\,*}(-p_\Omega) \, .
\ee


\end{document}